\documentclass[12pt,final]{article}
\usepackage{t1enc}
\usepackage[latin2]{inputenc}
\usepackage{showkeys}
\usepackage[dvips]{graphicx}
\usepackage[margin=2cm]{geometry}

\begin{document}

\title{\Large \bf The influence of line tension on the formation of liquid bridges in atomic force microscope-like geometry}
\author{F. Dutka and M. Napi\'orkowski\\ 
Instytut Fizyki Teoretycznej, Uniwersytet Warszawski,\\  00-681 Warszawa, Ho\.za 69, Poland}
\date{}
\maketitle{}
\abstract{The phase diagram of a fluid confined between a planar and a conical walls modelling the atomic force microscope 
geometry displays transition between two phases, one with a liquid bridge connecting the two walls of the microscope, and 
the other  without bridge. The structure of the corresponding coexistence line is determined and its 
dependence on the value of the line tension coefficient is discussed.}

\section{Introduction}

The behavior of fluids confined in nanosized structures depends sensitively on the properties of enclosing walls  \cite{Eijkel,Squires}. The equilibrium paradigm of such influence is provided by the capillary condensation phenomenon in a slit. Despite thermodynamic conditions favouring the gas phase in the bulk, the liquid phase may fill the space between the slit walls  \cite{Gennes}-\cite{Talanquer}. The shift of the gas-liquid coexistence line $\mu=\mu_{h}(T)$ in the slit with respect to its position in the bulk system $\mu=\mu_{0}(T)$  is described by the Kelvin law: $\mu_{0}(T) - \mu_{h}(T) \propto 1/h$, where $h$ denotes the distance between the walls  \cite{Rowlinson}. 

When the walls confining the fluid are non-planar and are like in the AFM geometry then one may expect formation of liquid bridge linking the 
opposite walls \cite{Andrienko}-\cite{Jang1} at thermodynamic conditions that favor gas phase in the bulk. The size and the shape of the liquid bridge depend on system's geometry as well as on the thermodynamic state specified, say, by temperature and chemical potential. There exists an additional attractive capillary force acting between the walls linked by the bridge which must be taken into account during the Atomic Force Microscope measurements \cite{Charlaix}-\cite{Lubarsky}. 
Formation of the liquid bridge is also exploited in dip-pen nanolithography \cite{Jang2}-\cite{Vettiger}. There the existence of 
the  bridge enables the flow of particular type of molecules from the tip of the AFM onto the planar substrate. The size of produced patterns depends on the geometry of the bridge which itself is controlled by the thermodynamic state of the system and its walls' geometry. These two examples already show that the knowledge of precise thermodynamic and geometrical conditions for the existence of liquid bridges together with their morphological properties is of crucial importance. 

In this paper we investigate the morphological phase transitions taking place in 3d fluid confined by the walls resembling the AFM geometry. They consist of a planar substrate and the surface of a cone modelling the microscope's tip. Our study is based on a macroscopic approach in the grand canonical ensemble with the purpose of constructing the relevant phase diagram and determining  the possible shapes of the liquid-like bridge spanned between the planar substrate and the conical tip. In particular, we discuss the influence of the line tension on the structure of the coexistence curve and examine the special cases related to the filling transition.

\section{Shape of meniscus}

\begin{figure}[htb]
 \begin{center}
  \includegraphics{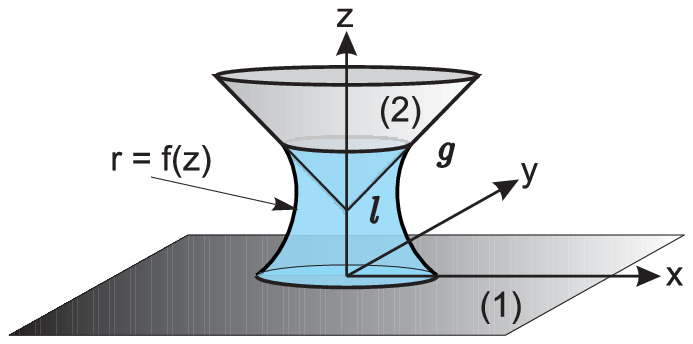}
 \caption{A liquid-like bridge ($l$) surrounded by the gas phase ($g$) connects the planar (1) and the conical (2) substrates. 
 \label{bridge} }
 \end{center}
\end{figure}

We consider fluid confined between two inert substrates, see Fig.\ref{bridge}. 
The surface of the lower substrate (1) is an infinite plane $z=0$ and the upper substrate (2) forms an infinite  
cone characterized by the opening angle $\pi-2\varphi$. The distance between the cone's apex and the plane $z=0$ is denoted by $h$; the 
parameters $h$ and $\varphi$ completely characterize the system's geometry which has cylindrical symmetry. In cylindrical coordinates the cone is described by equation $r=a(z)=(z-h)\cot \varphi$, where $r=\sqrt{x^2+y^2}$ and $z>h$. The system under study resembles the AFM-like geometry in which the conical substrate (2) plays the role of microscope's tip. In the limit $\varphi = 0$, $h \neq 0$ the geometry becomes that of a slit with two parallel walls separated by the distance $h$. In another limiting case corresponding to $\varphi \neq 0$ and $h=0$ the confining walls can be obtained by rotating a two-dimensional wedge with the opening angle $\varphi$ around the axis perpendicular to one of the walls and intersecting the wedge's apex. This confining geometry might allow - under appropriate thermodynamic conditions - for a filling transition \cite{Rejmer,Jakubczyk}. Thermodynamic states of the fluid are parametrized by temperature $T$ and chemical potential $\mu$, and are assumed to be away from the bulk liquid-gas critical point.  \\

Our macroscopic analysis is based on the grand canonical functional $\Omega([f], T, \mu, h, \varphi)$ which is parametrized  by the thermodynamic state $T, \mu$ and geometric parameters $h, \varphi$.  The allowed shapes of the liquid-gas interface $r=f(z)$ are  assumed to be cylindrically symmetric. The equilibrium shape $\bar f(z; T, \mu, h, \varphi)$ minimizes the functional $\Omega$ and leads to the thermodynamic grand canonical potential 
$\bar \Omega(T, \mu, h, \varphi) = \Omega([\bar f(z; T, \mu, h, \varphi)], T, \mu, h, \varphi)$. To shorten notation we shall refrain from displaying the dependence of analyzed functions and functionals on parameters $T, \mu, h, \varphi$ from now on. Comparison of the grand canonical potential values corresponding to configuration with and without the bridge allows us to tell which configuration is favorable under given thermodynamic and geometric conditions, and to construct the relevant phase diagram. 
In what follows we shall work with the functional $\Delta \Omega[f]$ representing the  difference between the grand potential corresponding to the state with a bridge and without it. For practical reasons, instead of temperature $T$ we switch to angles $\theta_{i}$, $i=1,2$ related to the temperature-dependent surface tension coefficients $\sigma_{ig},\sigma_{il}$, and $\sigma_{lg}$, $i=1,2$ via the Young equation \cite{Rowlinson} 
\begin{eqnarray}
\label{Young}
\sigma_{ig} - \sigma_{il} &=&  \sigma_{lg} \cos \theta_{i}, \quad i=1,2 \quad, 
\end{eqnarray} 
where $i$ stands for the $i$-th substrate. In what follows, the above relation will be applied also to thermodynamic states slightly off the bulk coexistence. \\
The functional $\Delta \Omega[f]$ contains the relevant surface free energies and the appropriate bulk contribution for states off the bulk coexistence. Contrary to the two-dimensional version of the this problem \cite{Dutka}, in the present case the line tension $\tau_{ilg}$  contributions related to the three phase contact lines where  gas, liquid and the $i$-th substrate meet should be taken into account. The grand canonical functional takes the following form
\begin{eqnarray} 
\label{delta_Omega}
 \frac{\Delta \Omega[f]}{2 \pi \sigma_{lg}} &=& \int {\rm d}z \left[ f(z) \sqrt{1+\left(\frac{{\rm d}f}{{\rm d}z} \right)^2}    
  + \frac{f(z)^2 - a(z)^2 \Theta(z-h)}{2 \lambda} \right]\Theta \Big(f(z) \Big) \, \Theta \Big(f(z)-a(z) \Big) \Theta(z)  
  \nonumber \\
  &&  + \int {\rm d}z \left[-\cos \theta_1 \,\frac{f(z)^2}{2} \delta(z) 
     - \frac{\cos \theta_2}{\sin \varphi} \, a(z) \, \Theta(z-h)\,\Theta \Big(f(z)-a(z) \Big) \right] \nonumber \\
  &&   +  \int {\rm d}z \,\Big[ \tilde \tau_{1lg} f(z) \delta(z)  +  \tilde \tau_{2lg} \cot \varphi \, 
  \Theta(z-h)\,\Theta(f-a)   \Big] \quad ,  
\end{eqnarray}
where parameters $\lambda = \sigma_{lg}/\Delta \mu \Delta \rho $ with $\Delta \mu =\mu_0-\mu \geq 0,  \Delta \rho =\rho_l-\rho_g > 0$, and $\tilde \tau_{ilg}=\tau_{ilg}/\sigma_{lg}$  have dimension of length. The symbols $\Theta(z)$ and $\delta(z)$ denote the Heaviside and Dirac function, respectively. 

The equilibrium interfacial shape $\bar f(z)$  minimizes $\Delta \Omega[f]$ and fulfills the equation
\begin{eqnarray} \label{shape}
\frac{1}{\bar f(z) \sqrt{1+\bar f'(z)^2}} - \frac{{\rm d}}{{\rm d}z} \frac{\bar f'(z)}{\sqrt{1+\bar f'(z)^2}} = - \frac{1}{\lambda}  
\end{eqnarray}  
supplemented by two boundary conditions
\begin{eqnarray} \label{boundaries}
 0 &=& \Bigg[ \cos \theta_1 + \frac{\bar f'(z)}{\sqrt{1+\bar f'(z)^2}} - \frac{\tilde \tau_{1lg}}{\bar f(z)}  \Bigg] \Bigg|_{z=0}  \nonumber  \\
 0 &=& \Bigg[\cos \theta_2 - \frac{\sin\varphi+\cos \varphi \bar f'(z)}{\sqrt{1+\bar f'(z)^2}} 
 - \frac{\tilde \tau_{2lg}}{\bar f(z)}\cos \varphi \Bigg] \Bigg|_{z=z_2} \quad ,   
\end{eqnarray}
where the coordinate $z_2$ is such that $\bar f(z_2)=a(z_2)$. The lhs of Eq.(\ref{shape}) contains the mean curvature \cite{Lazzer}  and thus the surface of the bridge has constant and negative mean curvature; it is called a concave nodoid \cite{nodoid1,nodoid2}. \\
Note that after introducing the contact angles $\gamma_1$ and $\gamma_2$ 
\begin{eqnarray} 
 \frac{{\rm d} \bar f}{{\rm d}z} \Bigg|_{z=0}   &=& - \frac{1}{\tan \gamma_1} \nonumber \\ 
 \frac{{\rm d} \bar f}{{\rm d}z} \Bigg|_{z=z_2} &=& \frac{1}{\tan(\gamma_2+\varphi)}   
\end{eqnarray} 
 Eqs. (\ref{boundaries}) take the form of modified Young equations \cite{Swain}
\begin{eqnarray} 
 \cos \gamma_1 &=& \cos \theta_1 - \frac{\tilde \tau_{1lg}}{\bar f(0)}  \label{gamma1} \\
 \cos \gamma_2 &=& \cos \theta_2 - \cos \varphi \,\frac{\tilde \tau_{2lg}}{\bar f(z_2)}  \label{gamma2} \quad .
\end{eqnarray} 

\begin{figure}[htb]
 \begin{center}
  \includegraphics{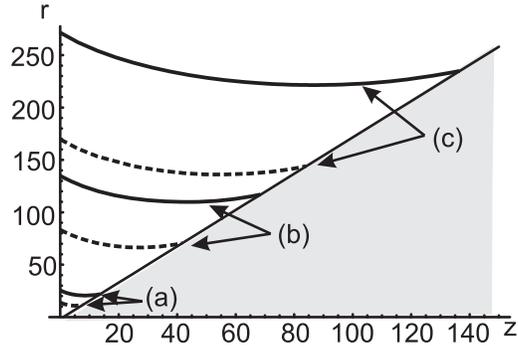}
 \caption{The shapes of bridges in two- (solid lines) and three-dimensional (broken lines) cases for specific choice of parameters 
$\varphi=\pi/6$, $\theta_1=\theta_2=\pi/6$, $\tilde \tau_{1lg} = \tilde \tau_{2lg}=0$, and (a) $\lambda = 10 h$, (b) $\lambda = 50 h$, (c) $\lambda = 100 h$. The variables $r$ and $z$ are displayed in $h$ units.   
 \label{ksztalt} }
 \end{center}
\end{figure}

\noindent Unfortunately Eq.(\ref{shape}) supplemented by the above boundary conditions, Eqs.(\ref{boundaries}), can not be solved analytically \cite{nona}. One has to resort to numerical procedures and we choose the shooting method \cite{Press}. For a given starting point $z_2>h$  the quantities $\bar f(z_2)$ and $\bar f'(z_2)$ are chosen in accordance with the boundary condition in Eq. (\ref{gamma2}), and the function $\bar f(z)$ is constructed towards the point $z=0$ by solving Eq. (\ref{shape}). Once the point $z=0$ is reached, the boundary condition in Eq. (\ref{gamma1}) is checked. If it's fulfilled  then the constructed function is accepted as $\bar f(z)$ (see Fig. \ref{ksztalt}); if it's not then the whole procedure must be repeated for a new choice of the starting point $z_2$. The existence of the solution of Eq.(\ref{shape}) depends on the choice of the thermodynamic state and the values of the geometric parameters. If there is no solution $\bar f(z)$ to Eq.(\ref{shape}) then the stable phase of our system corresponds to the absence of a liquid bridge. 
\mbox{Fig. \ref{ksztalt}} shows interfacial shapes (broken lines) in case of identical substrates, fixed distance $h$, zero line tension coefficients, and three different choices of angle $\varphi$; for reference, the corresponding 
shapes in two-dimensional case with the same choice of parameters are plotted (solid lines).

\section{Phase diagrams}

When the grand canonical potential difference is evaluated for solution $\bar f(z)$ of Eq.(\ref{shape}) then depending on 
the sign of $\Delta\Omega[\bar f]$ three cases are possible: (a) $\Delta \Omega [\bar f] < 0$ -- the bridge phase is favorable, (b) 
$\Delta \Omega [\bar f] > 0$ -- phase without bridge is favorable, (c) $\Delta \Omega [\bar f]=0$ -- two previous phases coexist  along a line in the ($T, \mu$)-plane which will be denoted as  $\mu = \mu_{AFM}(T)$. \\  

Before presenting the results for the currently studied three-dimensional case we recall that similar analysis performed for the 
two-dimensional case showed that both the Kelvin law for capillary condensation in a slit of width $h$ and the complete filling transition in a wedge (with the opening angle $\varphi$) can be treated as special cases of bridge formation \cite{Dutka}. In particular, the lowest temperature at which the bridge can be formed (for given angle $\varphi$) coincides with the wedge filling temperature  $T_f(\varphi)$ ($\theta(T_f(\varphi)) = \pi/2 - \varphi/2$) which is an increasing function of $\varphi$ \cite{Rejmer,Jakubczyk} (system's geometry constrains the range of angle $\varphi$ to $[0, \pi/2]$).  Upon crossing the coexistence curve the system exhibits discontinuous phase transition: a bridge with circular liquid-gas interface is formed whose width $\ell_h  \propto 1/\Delta\mu$ additionally depends on the opening angle $\varphi$ and becomes infinite for parallel walls, i.e.,  for $\varphi=0$. The dependence of the phase diagram on parameters $h$ and $\Delta\mu$  enters via the product $h\Delta\mu$. \\ 

\begin{figure}[htb]
 \begin{center}
  \includegraphics{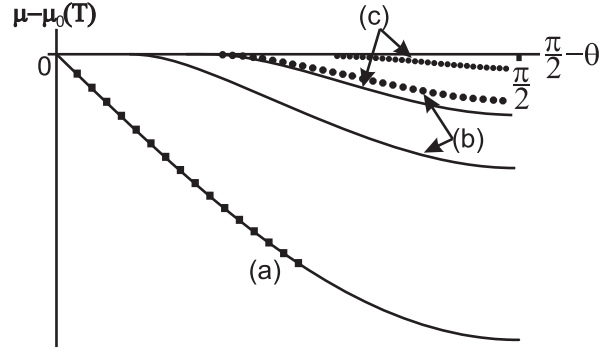}
 \caption{ The schematic phase diagram in variables $\mu - \mu_{0}$ ( measured in $\sigma_{lg}/\Delta \rho h$ units) and $\pi/2-\theta$ displaying the coexistence lines in three- (dotted lines) and two-dimensional system (solid lines) of phases in which the liquid bridge is present and absent, respectively,
  for the case of identical substrates ($\theta_1=\theta_2$), vanishing 
line tension coefficients ($\tilde \tau_{1lg}=\tilde \tau_{2lg}=0$), and for different values of the angle $\varphi$: 
(a) $\varphi = 0$,  (b) $\varphi = \pi/6$, (c) $\varphi = \pi/4$.  
 \label{2Dvs3D} }
 \end{center}
\end{figure}

\begin{figure}[htb]
 \begin{center}
  \includegraphics{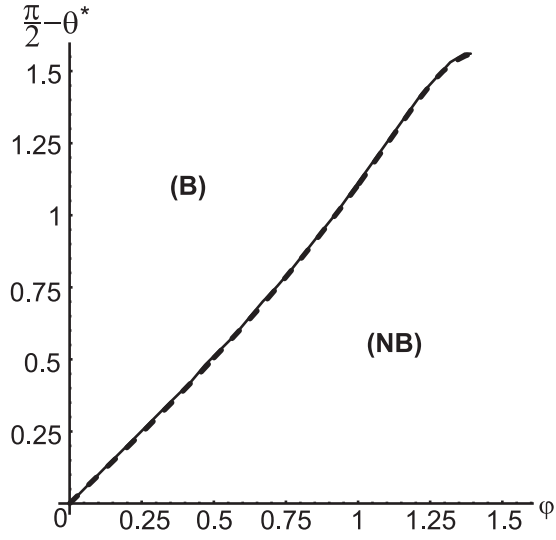}
 \caption{The value of angle $\pi/2-\theta^{*}$ at which the filling transition takes place as function of angle $\varphi$. The broken line represents the results of numerical analysis described in the text for the case $\tau=0$; the solid line is 
obtained under the assumption that catenoid describes the bridge's surface. 
 \label{stozek_plaszcz_theta_fi}}
 \end{center}
\end{figure}

The phase diagrams obtained for the three-dimensional system  in the case of vanishing line tensions, i.e.,  
for $\tilde \tau_{1lg}=\tilde \tau_{2lg}=0$ do not differ qualitatively from 
those in two dimensions, see Fig.\ref{2Dvs3D}. Both for two and three dimensions the coexistence line 
$\mu_{AFM}$ is a decreasing function of the angle $\pi/2-\theta$.  
Obviously, the coexistence lines in two- and three-dimensions are identical in the special case of a slit corresponding to 
$\varphi = 0$. Similarly as in two-dimensions, the filling transition temperature plays a distinguished role; the  AFM coexistence lines meet the bulk coexistence tangentially at the value of parameter $\pi/2 - \theta$ corresponding to the filling temperature $T_{f}(\varphi)$.  This particular value is denoted as $\pi/2 - \theta^{*}$, where $\theta^{*}=\theta(T_{f}(\varphi))$.  Fig.\ref{stozek_plaszcz_theta_fi} presents the numerically obtained plot of $\pi/2 - \theta^{*}$ as function of angle $\varphi$. 
It is worthwhile to note that the value $\theta^*=0$ is reached for $\varphi=\varphi_{0}<\pi/2$, where $\varphi_{0}=1.42$. Technically, $\varphi_{0}$  presents the largest value of angle $\varphi$ for which inscribing a catenoid (corresponding to angle $\theta = \theta^*$) into AFM-geometry is possible. 

The above results were obtained within macroscopic analysis which does not take into account the interaction between the liquid-gas interface and the substrates as specified e.g., by the 
interface potential $\omega(f)$ \cite{Dietrich, Schick}. We note that the formulation of this problem on the mesoscopic scale in which the  interface potential plays important role meets from the very outset some basic questions related to the structure of the  effective Hamiltonian relevant for the present geometry in which the substrate's curvature varies along the AFM tip. One possible way of dealing with these problems is to start the analysis from microscopic level as specified, e.g. by the density functional theory \cite{Evans2}-\cite{Koch}.

\section{The role of line tension}

\begin{figure}[htb]
\begin{center}
  \includegraphics{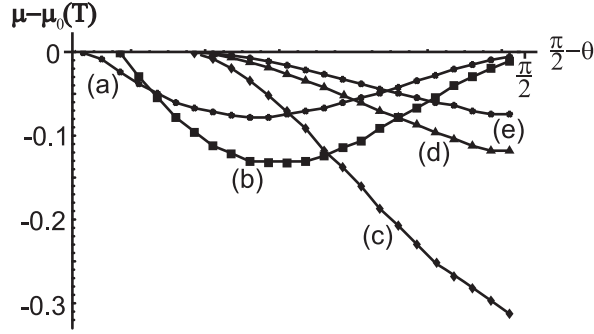}
 \caption{Coexistence lines corresponding to different values of line tension: (a) $\tilde \tau_{1lg} =\tilde\tau_{2lg} = -2 h$, 
(b) $ \tilde\tau_{1lg} =\tilde\tau_{2lg}= - h$, (c) $ \tilde\tau_{1lg} =\tilde\tau_{2lg}= 0$, (d) $\tilde\tau_{1lg} =\tilde\tau_{2lg}= h$, (e) $\tilde\tau_{1lg} =\tilde\tau_{2lg} = 2h$ 
for specific choice of $\varphi = \pi/6$. The variable $\mu - \mu_0(T)$ is measured in $\sigma_{lg}/\Delta \rho h$ units.
 \label{tau_diagrams} }
 \end{center}
\end{figure}

\begin{figure}[htb]
 \begin{center}
  \includegraphics{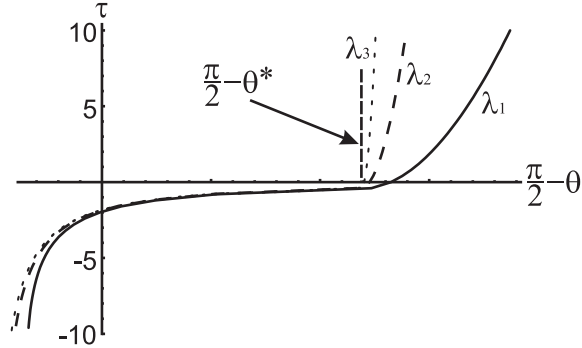}
 \caption{The values of line tension for which the corresponding coexistence line intersects the line  (a) 
$\lambda = \lambda_1=10^3 h$ (solid line), (b) $\lambda =\lambda_2=10^4 h$ (broken line), (c) $\lambda =\lambda_3=10^5 h$ 
(dotted line). For $\lambda \to \infty$ ($\mu = \mu_0$) 
the resulting plot becomes vertical at $\pi/2-\theta^*$ for positive $\tau$. The opening angle is fixed at $\varphi = \pi/6$.  
 \label{tau-theta} }
 \end{center}
\end{figure}

\begin{figure}[htb]
 \begin{center}
  \includegraphics{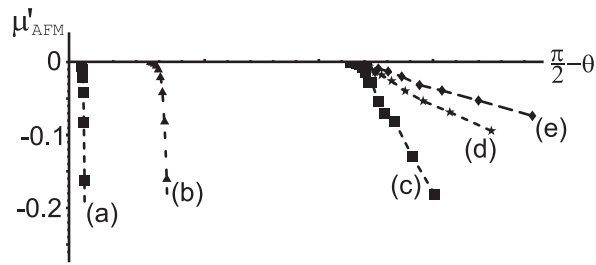}
 \caption{Numerical derivatives of the coexistence curve $\mu=\mu_{AFM}(\pi/2-\theta)$ with respect to its argument denoted as 
 $\mu^{,}_{AFM}$ for different line tensions ($\tilde \tau_{1lg} = \tilde \tau_{2lg} = \tilde \tau$): (a) $\tilde \tau = -2 h$,
 (b) $\tilde \tau = - h$,(a) $\tilde \tau = 0$,(d) $\tilde \tau = h$,(e) $\tilde \tau = 2h$ for specific choice 
 of $\varphi = \pi/6$ and $\theta_1=\theta_2 = \theta$.
 \label{pochodne} }
 \end{center}
\end{figure}

\begin{figure}[htb]
 \begin{center}
  \includegraphics{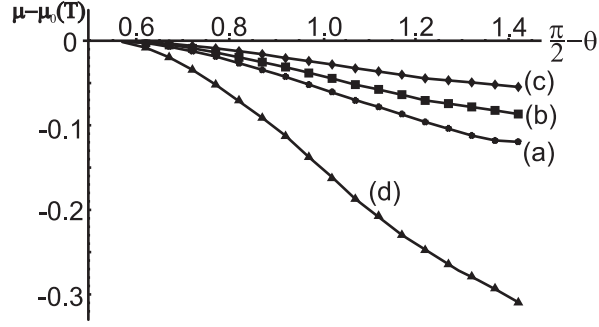}
 \caption{The coexistence lines corresponding to fixed $\tau = 1 a.u.$ and for different heights: 
(a) $h=1 a.u$, (b) $h=2 a.u.$, (c) $h=3 a.u.$. For $\tau=0$ the coexistence lines for different $h$ are identical, curve (d). 
The opening angle is fixed and equal $\varphi = \pi /6$. The difference $\mu - \mu_{0}(T)$ is measured in 
$\sigma_{lg}/\Delta \rho h$ units.   
 \label{skalowanie} }
 \end{center}
\end{figure}

In order to determine the influence of line tension on the phase diagram's structure one must know the line tension  
coefficient's dependence on the thermodynamic state of the system within the range of interest here. In spite of rather vast literature on this subject \cite{Getta}-\cite{Quere} there is still lack of general statements applicable outside immediate vicinity of the wetting points where universal behavior is observed.  
We recall that the line tension coefficient $\tau_{ilg}$ can be of either sign and its value changes significantly in the vicinity 
of the wetting temperature $T_{wi}$, \cite{Indekeu1}-\cite{Wang}. The behaviour of the line tension coefficient near $T_{wi}$ depends on the order of the wetting transition and on the type of interaction between particles constituting the system. The universal property of this dependence is that the line tension coefficient is  an increasing function for 
$T \nearrow T_{wi}$, and the limiting value $\tau_{ilg}(T_w)$ is positive for the first-order wetting (can be even infinite 
for specific choice of interactions), and is zero for critical wetting \cite{Indekeu1}. \\
In view of the lack of precise information on the dependence of line tension coefficients on chemical potential and temperature 
we shall estimate their influence on the phase diagram by assuming them to be constant parameters. We shall examine the system 
properties corresponding to different constant values and, in particular, both signs will be taken into account. For simplicity, in the following  analysis we again consider identical substrates, i.e., 
$\theta_1=\theta_2 = \theta$ and $\tilde \tau_{1lg} = \tilde \tau_{2lg} = \tau$. \\ 
The line tension influences the grand canonical potential not only through its presence in Eq.(\ref{delta_Omega}) 
but  also via 
boundary conditions where it affects the contact angles, Eqs(\ref{gamma1},\ref{gamma2}). The coexistence lines corresponding 
to different values of the line tension are shown on Fig.\ref{tau_diagrams}. For positive $\tau$-values the shape of 
the coexistence line does not change qualitatively with respect 
to the $\tau=0$ case; it is still monotonic, i.e., $\mu_{AFM}$ is an increasing function of $\theta$ and achieves the lowest 
value at the wetting temperature where $\theta (T=T_w)=0$. \\On the other hand, for $\tau<0$  the coexistence lines are not 
monotonic; they exhibit a local minimum and approach the bulk coexistence line  $\mu=\mu_0$ for $T \to T_w$, 
see Fig.\ref{tau_diagrams}.  
One can argue for this type of behaviour at small $\theta$ values in the following way based on Eqs(\ref{gamma1}, \ref{gamma2}): close to the wetting temperature both $\bar f(z_2)$ and $\bar f(0)$ must  increase in order to keep  
$\cos \gamma_1, \cos \gamma_2 <1$. Thus the size of the bridge which is proportional to the $(\mu_0 - \mu)^{-1}$ also 
increases which means that $\mu_{AFM}$ approaches $\mu_0$. 
\\
The presence of line tension also influences the temperature at which the coexistence lines approach (tangentially)  
the bulk coexistence. For positive line tensions this particular temperature does not depend on the actual $\tau$-value. 
For negative line tensions this temperature decreases with $\mid \tau \mid$. This behaviour is illustrated on Figs \ref{tau-theta},  \ref{pochodne}.
\\
For two-dimensional systems \cite{Dutka} and for three-dimensional systems with $\tau = 0$ there is only one quantity with dimension of length which can be built out of the system  parameters, namely $\lambda = \sigma_{lg}/\Delta \mu \Delta \rho$. The equation for the coexistence line (corresponding to $\Delta \Omega [\bar f] = 0$) can be rewritten as 
 ${\cal F}_0 \left( \theta, \frac{\lambda}{h} \right) = 0 $,    
where ${\cal F}_0$ is a dimensionless function. Thus the phase diagrams displayed in variables $\pi/2-\theta$ and 
$ \mu - \mu_0(T)$ measured in $\sigma_{lg}/\Delta \rho h$ do not depend any more on height $h$. If one considers system with  
non-zero line tension then there is one additional parameter with the dimension of length, namely $\tilde \tau$. This time the 
equation for the coexistence curve  takes the form ${\cal F}_\tau \left(\theta, \frac{\lambda}{h} , \frac{\tilde\tau}{h}  \right) = 0 $  
and the coexistence lines corresponding to different $h$-values are not identical unless $\tilde \tau$ is properly rescaled, see  Fig.\ref{skalowanie}. 

\section{Sizes of liquid bridges}

The crucial aspect of the present analysis is that it deals with macroscopic objects. 
In Fig.\ref{ksztalt} we see that the shape of the bridges's surface depends on the distance from the bulk coexistence measured by 
$\mu-\mu_0$. In this paragraph we would like to analyze the size of the liquid bridge in more detail, and to see how 
it depends on parameters describing the system. 

\begin{figure}[htb]
 \begin{center}
  \includegraphics{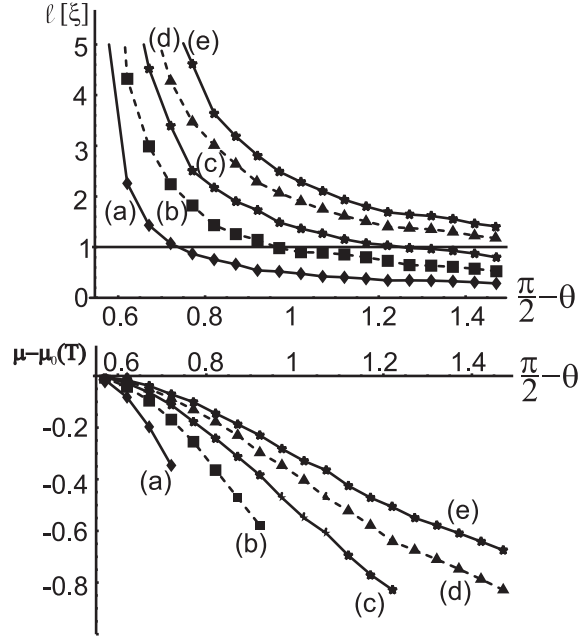}
 \caption{The bridge's width $\ell$ in $\xi $ units (upper graph) and the corresponding phase diagram with the coexistence curves (lower graph) displayed  for parameter values at which macroscopic approach is applicable, i.e., for $\ell \geq 1$  for different $h$-values: (a) $h=0.1 \xi$, (b) $h=0.3 \xi$, (c) $h=0.5 \xi$, (d) $h=0.7 \xi$, (e) $h=0.9 \xi$. The angle $\varphi = \pi/6$ and $\tau =0$. The difference $\mu - \mu_{0}(T)$ is measured in $\sigma_{lg}/\Delta \rho \xi$ units.  
 \label{width_h} }
 \end{center}
\end{figure}

\begin{figure}[htb]
 \begin{center}
  \includegraphics{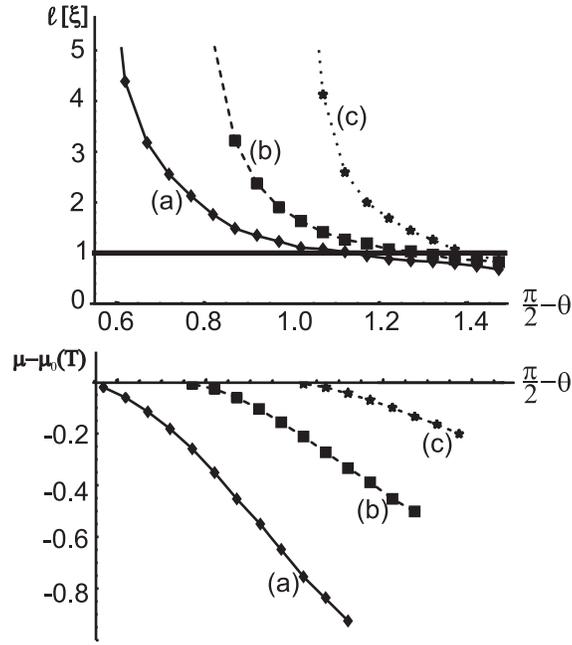}
 \caption{The bridge's width $\ell$ in $\xi $ units (upper graph) and the corresponding coexistence curves (lower graph) displayed only for parameter values at which macroscopic approach is applicable for different angles: (a) $\varphi=0.5$, (b) $\varphi=0.7$ and (c) $\varphi=0.9$. The height $h = 0.25 \xi$ and $\tau =0$. The difference $\mu - \mu_{0}(T)$ is measured in $\sigma_{lg}/\Delta \rho \xi$ units.  
 \label{width_fi} }
 \end{center}
\end{figure}

\begin{figure}[htb]
 \begin{center}
  \includegraphics{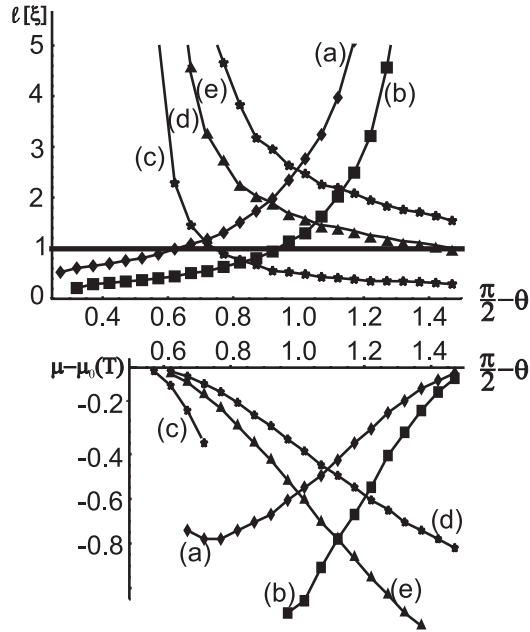}
 \caption{The bridge's width $\ell$ in $\xi $ units (upper graph) and the corresponding coexistence  curves (lower graph) displayed for parameter values at which macroscopic approach is applicable (b) for different values of line tension: (a) $\tau = -0.2 \xi$, (b) $\tau = -0.1 \xi$, (c) $\tau = 0$, (d) $\tau = 0.1 \xi$, (e) $\tau = 0.2 \xi$. The height $h = 0.1 \xi$ and $\varphi = \pi/6$. 
The difference $\mu - \mu_{0}(T)$ is measured in $\sigma_{lg}/\Delta \rho \xi$ units. 
 \label{width_tau} }
 \end{center}
\end{figure}

As the measure of the width $\ell$ of the bridge we take the minimal value of $\bar f(z)$ in the range of $z=0$ to $z=z_2$.  
In the case of macroscopic bridges considered in this paper their width must be large compared to the bulk correlation length 
$\xi$ which we take to be of the order  $\xi \approx 0.1 nm$. 
The way the width $\ell$ depends on height $h$, angle $\varphi$, and the line tension are depicted on Figs \ref{width_h}-\ref{width_tau}. For temperatures $\theta \leq \theta^*(\varphi)$ the width of the bridge grows to infinity upon approaching the bulk coexistence whereas for $\theta > \theta^*(\varphi)$ the width of the bridge achieves a fixed value for $\mu \to \mu_0(T)$. Obviously, the latter situation takes place only for negative line tension coefficients, $\tau<0$.

For small $h$-values  and  $\tilde\tau \approx 0.1\xi$ (for typical values of 
$\tau = 10^{-11} J/m$ and $\sigma_{lg} = 10^{-2} J/m^2$ the coefficients $\tilde\tau = \tau/\sigma_{lg}$  are in the nanometer range) our macroscopic analysis predicts  for $\varphi$ close to $\pi/2$ 
the existence of bridges with width smaller than $\xi$. One has to treat these results with caution because such situation requires 
more detailed approach than the present macroscopic description, see \cite{Hauge} for similar arguments in the case of wedge filling 
transitions. We note that the minimal size of the bridge included into our analysis can not be smaller then the largest among the 
parameters $h$ and $\tilde \tau$ which have the dimension of length. This restriction reduces the range of parameters 
for which reliable phase diagrams can be constructed within present approach.

\section{Summary}

We analyzed macroscopically the formation of liquid bridges in AFM-like geometry in three dimensions with the microscope tip modeled by cone. 
In the case when the line tension is not taken into account the properties of the phase diagram displaying the coexistence of two phases, one with the liquid bridge present and the other without bridge, is similar to the phase diagram obtained for two-dimensional version of this problem. The coexistence line is an increasing function of the contact angle $\theta$, and it meets the bulk coexistence line tangentially at the value of the contact angle $\theta^{*}(\varphi)$ corresponding to the filling transition. We provide numerically obtained plot of $\pi/2 - \theta^{*}$ versus $\varphi$ which points - at least for small $\varphi$ values - at linear dependence with prefactor $1$. We also note that the particular value $\theta^{*}=0$ corresponding to wetting is obtained not for $\varphi=\pi/2$ but for smaller value $\varphi=\varphi_{0}=1.42$. \\    
In the presence of non zero line tension values, in particular negative ones, the shape of the coexistence line changes substantially. For negative values of line tension coefficients the coexistence lines exhibit local minima and reapproach   
the bulk coexistence ($\mu = \mu_0$) for temperatures close to $T_{wi}$. In addition we analyzed the size of the bridge present in the system and displayed its dependence on system parameters. 
Our investigations are limited to macroscopic bridges and to situations in which line tensions are treated as constants, 
i.e., do not depend on the thermodynamic state of the system. The requirement that the characteristic lengths of this problem are larger than the bulk correlation length reduces the range of parameters at which reliable conclusions can be drawn; this is depicted on Figs 9-11. \\ A more detailed analysis based on the capillary 
Hamiltonian \cite{Rejmer}, \cite{Jakubczyk} might allow one to estimate the role of fluctuations on the formation and stability of 
liquid bridges investigated in this paper. We also point at difficulties related to such mesoscopic formulation of the bridge formation problem.  \\

\noindent{\bf{Acknowledgement}}\\The authors express their gratitude to S. Dietrich, P. Jakubczyk, and A. Majhofer for helpful discussions. 

This work has been financed from resources provided for scientific research for years 2006-2008 as a research project 
N202 076 31/0108.

\newpage



\begin{thebibliography}{99}
\bibitem{nona} After change of variables $u(\bar f)=\frac{{\rm d} \bar f}{{\rm d} z}(\bar f)$ Eq. (\ref{shape}) takes the following form 
\begin{eqnarray}
\frac{{\rm d} u^2}{{\rm d} \bar f} &=& \frac{2}{\lambda}(1+u^2)^{3/2}+\frac{2}{\bar f}(1+u^2) \quad ,
\end{eqnarray}
and can be rewritten as 
\begin{eqnarray} \label{elliptical}
 \int {\rm d}\bar f \ z(\bar f) &=&  \pm \int \frac{\bar f^2-2C}{\sqrt{(2\lambda \bar f)^2-(\bar f^2-2C)^2}} {\rm d}\bar f \quad ,
\end{eqnarray}
where $C$ is the integration constant. The elliptical integral on the rhs cannot be calculated analytically. The function $\bar f(z)$ cannot be presented in algebraic form as for the two dimensional problem \cite{Dutka}.
\end{thebibliography}
\end{document}